**Floating up of the zero-energy Landau level in monolayer epitaxial graphene**

Lung-I Huang[1,2], Yanfei Yang[1,3], Randolph E. Elmquist[1],
Shun-Tsung Lo[4, *], Fan-Hung Liu[4], and Chi-Te Liang[2,4, *]

[1]*National Institute of Standards and Technology (NIST), Gaithersburg, MD 20899, USA*
[2]*Department of Physics, National Taiwan University, Taipei 106, Taiwan*
[3]*Department of Physics, Georgetown University, Washington, DC 20057, USA*
[4]*Graduate Institute of Applied Physics, National Taiwan University, Taipei 106, Taiwan*

We report on magneto-transport measurements on low-density, large-area monolayer epitaxial graphene devices grown on SiC. We show that the zero-energy Landau level (LL) in monolayer graphene, which is predicted to be magnetic field ($B$)-independent, can float up above the Fermi energy at low $B$. This is supported by the temperature ($T$)-driven flow diagram approximated by the semi-circle law *as well as* the $T$-independent point in the Hall conductivity $\sigma_{xy}$ near $e^2/h$. Our experimental data are in sharp contrast to conventional understanding of the zeroth LL and metallic-like behavior in pristine graphene prepared by mechanical exfoliation at low $T$. This surprising result can be ascribed to substrate-induced sublattice symmetry breaking which splits the degeneracy of the zeroth Landau level. Our finding provides a unified picture regarding the metallic behavior in pristine graphene prepared by mechanical exfoliation, and the insulating behavior and the insulator-quantum Hall transition in monolayer epitaxial graphene.

When a strong magnetic field $B$ is applied perpendicular to the plane of monolayer graphene,[1-3] Landau quantization results in a series of Landau levels whose energy is given by[4]

$$E_N = \text{sgn}(N)\sqrt{2\hbar v_F^2 eB|N|}, \qquad (1)$$

where $N$, $\hbar$, $e$, $v_F$ are an integer, reduced Planck constant, electronic charge and Fermi velocity, respectively. According to Eq. (1), the energy of the $N = 0$ Landau level (LL) is zero and thus is independent of $B$. Such a zeroth LL, which is shared equally by electrons and holes with degeneracy of four, is unique in graphene and has no counterparts in any semiconductor-based two-dimensional (2D) systems.

Although in most cases, transport in pristine graphene (PG) on $SiO_2$ prepared by mechanical exfoliation shows metallic behavior or a very weak $T$ dependence,[1, 2] insulating behavior may appear when sublattice symmetry is broken.[5] Interestingly, recent experiments show very low conductivity near the charge neutrality point for monolayer graphene on boron nitride with a suspended top gate[5] and for monolayer epitaxial graphene (EG) with a point-like constriction caused by bilayer patches[6]. Such important results on monolayer graphene suggest further studies are required and may be related to the possible splitting of the zeroth LL (Ref. 7) at low $B$. Here, we address the two aforementioned fundamental issues: the fate of the zero-energy LL at low fields and the insulating behavior in disordered graphene. We shall show that in strongly disordered EG, the $N = 0$ electron LL can *float up* above the Fermi energy $E_F$ at low $B$ as evidenced by a well-defined $T$-independent point in the measured Hall conductivity $\sigma_{xy}$ and the appearance of a semicircle relation in the $T$-driven flow diagram.[8] Our new results are in sharp contrast to the conventional understanding of the zero-energy LL which is believed to be *B*-independent. Moreover, our data provide a thorough understanding of the low-field insulator-quantum Hall (I-QH) transition in disordered EG as well as the metallic-like behavior in PG.

Our EG devices were fabricated utilizing a clean lithography process[9] that leaves the surface free of resist residues. After the fabrication process, doping occurs due to or initiated by chemical etching of the protective layer and exposure to air. We have engineered the carrier density as low as $n \approx 10^{15}$ m$^{-2}$. Here, the exposed Si atoms in the SiC (0001) lattice form partial covalent bonds to carbon atoms in the lower graphene layer (buffer layer), and only the top layer is conducting. Si-C covalent bonds and defects such as interfacial dangling bonds affect the electrical environment of the graphene sheet and can break its sublattice symmetry.[10] Low carrier density is known to reduce the screening of Coulomb potential fluctuations, and therefore enhances the substrate influences on the conducting sheet being sustained.

Large-area EG devices are suitable for studies of QH transitions and insulating behavior since the long-range effects of increasing disorder may be hidden by local or size-dependent phenomena for small samples.[11] Moreover, in EG grown on SiC,[12, 13] $E_F$ can be pinned to the localized states[14] such that the $v = 2$ QH plateau extends from a low field (~ 1 T) to exceptionally high values (30 T),[15] making EG an ideal system for studying an isolated low-field QH transition, although no such high-field transition has been reported. A possible reason for this is the reservoir model responsible for the long $v = 2$ QH plateau[14] so that one does not observe the high-field insulating state. Measurements on large-area (0.6 mm × 0.1 mm) devices were made in a perpendicular magnetic field up to 9 T in a variable-temperature cryostat using standard low-frequency lock-in techniques.

The longitudinal and Hall resistivities ($\rho_{xx}$ and $\rho_{xy}$) for the three samples (EG1, EG2, and EG3) at various $T$ are plotted in Figs. 1a-c. Since the low-$T$ resistivity of EG1 is more than ten times lower than that of EG3 and nearly two times lower than that of EG2, we describe EG1 (EG3) as the least (most) disordered device. We can immediately see the $T$-independent points in $\rho_{xx}$ at crossing fields $B_c$ in all three samples. For $B < B_c$, the device behaves as an insulator in the sense that $\rho_{xx}$ decreases with increasing $T$.[16] For $B > B_c$, the device shows QH-like behavior and $\rho_{xx}$ increases with increasing $T$.[16-19] Our results show characteristics of the insulator to $\nu = 2$ QH transition observed in disordered 2D systems.[17-19] However, the anomalous $\nu = 2$ quantum Hall plateau in $n$-type monolayer graphene[2,3] is observed when the electrons fill *both* the electron and hole-like zeroth LL with total degeneracy of 4. This is in sharp contrast to the $\nu = 2$ QH plateau observed in conventional semiconductor-based disordered 2D electron system[17-19] when the electron only fills the spin-degenerate zeroth LL (degeneracy = 2). Like other disordered 2D systems, localization and interaction effects are observed in our devices (see Supporting Information).

To further study the observed I-QH transition, we plot $\sigma_{xx}$ and $\sigma_{xy}$ for EG1, EG2 and EG3 in Figs. 2a-c. Interestingly, a clear $T$-independent crossing point in $\sigma_{xy}$ develops near $e^2/h$ for EG2 and EG3. In the scaling theory of the QH effect, values of $\sigma_{xy}$ that are half multiples of $e^2/h$ (per spin) behave as unstable points under renormalization.[18] Therefore the observed crossing point near $e^2/h$ suggests a delocalization/localization process occurs when the zeroth LL passes *upwards* through $E_F$ when $B$ is decreased.[18]

A $T$-driven flow diagram in the ($\sigma_{xy}$, $\sigma_{xx}$) plane can be used to study the physics of localization processes in 2D systems.[20, 21] A field-induced transition involves a transition between two fixed points in this diagram, with a sudden increase and a similar decrease in $\sigma_{xx}$ once the LL is emptied or filled. It has been experimentally verified that this transition traces out a semicircle[20] in the ($\sigma_{xy}$, $\sigma_{xx}$) plane and for systems with a single conduction channel the semicircle represents a critical boundary for the QH state. The semicircle is centered at (0, $e^2/h$) and follows $(\sigma_{xx})^2 + (\sigma_{xy} - e^2/h)^2 = (e^2/h)^2$, where the transition to the $v = 2$ QH state occurs.

Figures 3a and 3b show that samples EG1 and EG2 develop robust $v = 2$ QH characteristics to the right of the semicircle ($\sigma_{xy} > e^2/h$) at fields $B \approx 1$ T, and approach the limiting point of the QH state at ($2e^2/h$, 0). Conductivity data is given in Fig. 3 for all three EG samples with arrows showing $T$-driven flow superimposed at a series of fixed $B$. For a given sample, results at constant magnetic field strength that follow a vertical $T$-driven flow line corresponds to a critical field denoted by $B_c^\sigma$ identified as a crossing point of constant conductivity $\sigma_{xy}$. Similar curved arrows show how flow divides along the critical boundary of the QH state, shown by a dotted semicircle, starting at an unstable point indicated by a black dot. The sample with the most pristine behavior (EG1) avoids the critical boundary with high conductivity ($\sigma_{xx} \approx 4e^2/h$) at low fields, and the vertical flow line occurs at $\sigma_{xy} < e^2/h$. Vertical $T$-driven flow arrows in Figs. 3b and 3c show that the crossing magnetic field $B_c^\sigma$ occurs close to $\sigma_{xy} = e^2/h$ for both EG2 and EG3, while the magnitude of $\sigma_{xx}$ decreases from $\sigma_{xx} \approx 2e^2/h$ to $\sigma_{xx} \approx e^2/h$. Thus, we can characterize the $T$-driven flow for increasing disorder strength in our samples by vertical flow along $\sigma_{xy} = e^2/h$, the line that points toward the center of the $v = 2$ QH semicircle. Elsewhere the flow diverges from verticality especially near the semi-circle boundary, as clearly seen for sample EG3, where flow lines become nearly tangent to the semicircle.

Based on the floating up picture,[22, 23] Kivelson, Lee, and Zhang have proposed the global phase diagram (GPD) which describes possible phase transitions in a 2D system.[24] When the spin degeneracy is considered, for a strongly disordered 2D system in which the spin-splitting is not well-resolved, the only I-QH transition is the 0-2 transition, where the numbers 0 and 2 correspond to the insulating phase and the $v = 2$ QH state. This 0-2 transition and the 2-0 transition, from the QH state to the insulating regime, are equivalent within the GPD framework.[24] The establishment of the semicircle relation for the 0-2 transition requires that the lowest extended band continuously floats up above $E_F$ with smaller $B$.[17, 18, 22-24] The semicircle-like flow lines obtained on EG3 therefore provide compelling evidence for the levitation of the zeroth LL in disordered graphene, linking the observed insulating behavior in EG3 at low fields to the zeroth Landau band floating up above $E_F$. The semicircle law fails to provide a good explanation for the transition in the cleaner devices EG1 and EG2. The possible origin is that their weak disorder prohibits the observation of the levitating Landau band. Moreover we found that the slope of $\sigma_{xy}$ at $B_c^\sigma$ scales with temperature following $T^{-\kappa}$ with $\kappa = 0.21$ and 0.36 for EG2 and EG3, respectively (see Fig. S7 in the Supporting Information). At such low-field transitions, the Zeeman splitting plays a minor role, preserving the spin degeneracy. Therefore the increase in $\kappa$ is attributed to the breaking of sublattice symmetry in the presence of potential fluctuations, which splits the zeroth Landau band.[25] With increasing disorder from EG1 to EG3, the semicircle relation between $\sigma_{xy}$ and $\sigma_{xx}$ becomes apparent, due to the underlying levitation of the zeroth Landau band.

For EG3, the *T*-independent crossing point in $\rho_{xx}$ occurs at the filling factor $\nu_c = nh/(eB_c)$ of 0.6, which is in agreement with the recently reported value for the high-field levitation of the zeroth Landau band.[26] However for EG1 and EG2, it corresponds to $\nu_c$ = 16 and $\nu_c$ = 7, respectively, which is much higher than that for EG3. These values deviate from the prediction of plateau-to-plateau transition between the $\nu = 6$ and $\nu = 2$ QH state, suggesting that the transition in weakly disordered EG1 and EG2 does not result from the $N = 1$ Landau band passing through the Fermi energy with magnetic field. In addition, we have estimated the width $\Gamma \approx \hbar/\tau$ of Landau level broadening due to disorder. The results are 23 meV, 24 meV, 76 meV for EG1, EG2, and EG3, respectively. However the Fermi energy lies at $E_F$ = 49 meV, 35 meV, and 28 meV for EG1, EG2, and EG3. Interestingly, for EG3, $E_F$ is smaller than the estimated $\Gamma$. This finding infers a narrowing of the zeroth Landau band, which is robust against some sorts of disorder[27] such that we can still observe the $\nu = 2$ quantum Hall character in highly disordered EG3. It is worthwhile noting that the finite size effect and the charge transfer from the buffer layer/SiC interface (which partially determines the carrier density in a QH state) to the graphene sheet[24] would modify the transitions.

Figure 4 shows schematically the main finding in this paper. Due to the graphene-substrate coupling, sublattice symmetry breaking splits the $N = 0$ Landau level into the $N = 0$ electron LL and the $N=0$ hole LL. When electrons fill the $N = 0$ hole LL, this filled LL does not contribute to the conduction therefore we only need to consider the $N = 0$ electron LL. At high *B*, where edge-state transport occurs, the zeroth electron LL is below $E_F$ and the device is in the QH regime. With decreasing *B*, the zeroth electron LL floats up, and there are no extended states below $E_F$ so that EG shows insulating behavior. Such results are in sharp contrast to transport in PG in which the zeroth LL is independent of *B*. Figure 4 is consistent with the fact that in most cases, PG on $SiO_2$ prepared by mechanical exfoliation shows metallic behavior in the sense that the resistivity increases with increasing $T$ or shows a very weak $T$ dependence.[1,2] Since the zeroth LL is $B$-independent and always zero and below $E_F$, strongly insulating behavior usually is not observed in PG. Thus there is no low-field I-QH transition in PG.

It was shown that graphene-substrate induced sublattice symmetry breaking coupled with charge disorder in epitaxial graphene layer can substantially modify the trasnport properties of graphene.[28] We note that in PG on h-BN, strongly insulating behavior solely due to graphene-substrate related sublattice symmetry breaking is observed.[5] Interestingly, such an insulating phase makes a direct transition to the $\nu = 0$ state at an extremely low field ($B \sim 0.1$ T) without an intermediate transition to the $\nu = 2$ QH state, in sharp contrast to our experiment.[5] In our case, the mobility of EG3 is 20 times lower than that of the PG on h-BN. The stronger disorder and the fact that our device is not exactly at the Dirac point should inhibit the formation of the $\nu = 0$ state as supported by no sign of the $\nu = 0$ plateau in $\sigma_{xy}$. Therefore although insulating behaviour can be observed in both PG on h-BN and disordered EG, we observe a transition from the insulating phase to the $\nu = 2$ QH state *as well as* the semi-circle-like $T$-driven flow diagram, evidence for floating up of the $N = 0$ electron LL due to stronger disorder compared with that of Amet *et al.*[5] Our results, together with the pioneering work of Amet *et al.* suggest that sublattice symmetry breaking plays an important role in the observed insulating behaviour in graphene subject to the environment effect. The strength of disorder, however, determines the allowable transition between the insulating state and the $\nu = 2$ QH state or the $\nu = 0$ state.

In conclusion, we have reported magneto-transport measurements on low-density monolayer EG with various amount of disorder. $T$-independent crossing points are observed in all three samples. We have found that the observed $T$-independent point in $\rho_{xx}$ survives after subtraction of the electron-electron interaction corrections (see Supporting Information), demonstrating that such crossing points are related to magnetic-field-induced delocalization/localization transitions. With increasing disorder, $T$-independent points in $\sigma_{xy}$ emerge, corresponding to the unstable points under renormalization in the scaling theory of the QH effect. Our results therefore suggest that $\sigma_{xy}$, rather than $\rho_{xx}$, is the more important physical quantity in the study of quantum Hall transitions. Most importantly, in the most disordered device, we have observed $T$-driven flow lines approximated by the semi-circle law. Such results provide compelling evidence that the zeroth LL is levitated at low $B$ and can explain the insulating behavior in our EG which is not normally observed in PG. Our experimental results shed light on the fate of the zeroth LL and the ground state of graphene near the Dirac point and at low magnetic fields.

ASSOCIATED CONTENT
Supporting Information
Sample preparation; analysis of quantum corrections due to localization and interactions; clarification of the low-field insulating behavior by removing the interaction corrections; physical quantities of the studied three devices

AUTHOR INFORMATION
Corresponding Authors
*E-mail: shuntsunglo@mail.ncku.edu.tw
*E-mail: ctliang@phys.ntu.edu.tw

Notes
The authors declare no competing financial interest.

ACKNOWLEDGMENTS

This work was funded by the Ministry of Science and Technology (MOST), Taiwan. C.T.L. was supported by the MOST, Taiwan (grant numbers MOST 103-2918-I-002-028, MOST 103-2622-E-002 -031, MOST 104-2622-8-002 -003 and MOST 102-2119-M-002 -016 -MY3).

REFERENCES

1. Novoselov, K. S.; Geim, A. K.; Morozov, S. V.; Jiang, D.; Zhang, Y.; Dubonos, S. V.; Grigorieva, I. V.; Firsov, A. A. *Science* **2004,** *306*, 666-669.
2. Novoselov, K. S.; Geim, A. K.; Morozov, S. V.; Jiang, D.; Katsnelson, M. I.; Grigorieva, I. V.; Dubonos, S. V.; Firsov, A. A. *Nature* **2005,** *438*, 197-200.
3. Zhang, Y.; Tan, Y.-W.; Stormer, H. L.; Kim, P. *Nature* **2005,** *438*, 201-204.
4. Castro Neto, A. H.; Guinea, F.; Peres, N. M. R.; Novoselov, K. S.; Geim, A. K. *Rev. Mod. Phys.* **2009,** *81*, 109-162.
5. Amet, F.; Williams, J. R.; Watanabe, K.; Taniguchi, T.; Goldhaber-Gordon, D. *Phys. Rev. Lett.* **2013,** *110*, 216601.
6. Chua, C.; Connolly, M.; Lartsev, A.; Yager, T.; Lara-Avila, S.; Kubatkin, S.; Kopylov, S.; Fal'ko, V.; Yakimova, R.; Pearce, R.; Janssen, T. J. B. M.; Tzalenchuk, A.; Smith, C. G. *Nano Lett.* **2014,** *14*, 3369-3373.
7. Giesbers, A. J. M.; Ponomarenko, L. A.; Novoselov, K. S.; Geim, A. K.; Katsnelson, M. I.; Maan, J. C.; Zeitler, U. *Phys. Rev. B* **2009,** *80*, 201403.
8. Hilke, M.; Shahar, D.; Song, S. H.; Tsui, D. C.; Xie, Y. H.; Shayegan, M. *Europhys. Lett.* **1999,** *46*, 775.
9. Yang, Y.; Huang, L.-I.; Fukuyama, Y.; Liu, F.-H.; Real, M. A.; Barbara, P.; Liang, C.-T.; Newell, D. B.; Elmquist, R. E. *Small* **2015,** *11*, 90-95.
10. Zhou, S. Y.; Gweon, G. H.; Fedorov, A. V.; First, P. N.; de Heer, W. A.; Lee, D. H.; Guinea, F.; Castro Neto, A. H.; Lanzara, A. *Nat. Mater.* **2007,** *6*, 770-775.
11. Nakajima, T.; Ueda, T.; Komiyama, S. *J. Phys. Soc. Jpn.* **2007,** *76*, 094703.
12. Berger, C.; Song, Z.; Li, T.; Li, X.; Ogbazghi, A. Y.; Feng, R.; Dai, Z.; Marchenkov, A. N.; Conrad, E. H.; First, P. N.; de Heer, W. A. *J. Phys. Chem. B* **2004,** *108*, 19912-19916.
13. Riedl, C.; Coletti, C.; Starke, U. *J. Phys. D: Appl. Phys.* **2010,** *43*, 374009.
14. Janssen, T. J. B. M.; Tzalenchuk, A.; Yakimova, R.; Kubatkin, S.; Lara-Avila, S.; Kopylov, S.; Fal'ko, V. I. *Phys. Rev. B* **2011,** *83*, 233402.
15. Alexander-Webber, J. A.; Baker, A. M. R.; Janssen, T. J. B. M.; Tzalenchuk, A.; Lara-Avila, S.; Kubatkin, S.; Yakimova, R.; Piot, B. A.; Maude, D. K.; Nicholas, R. J. *Phys. Rev. Lett.* **2013,** *111*, 096601.
16. Song, S. H.; Shahar, D.; Tsui, D. C.; Xie, Y. H.; Monroe, D. *Phys. Rev. Lett.* **1997,**

*78*, 2200-2203.

17. Jiang, H. W.; Johnson, C. E.; Wang, K. L.; Hannahs, S. T. *Phys. Rev. Lett.* **1993,** *71*, 1439-1442.

18. Hughes, R. J. F.; Nicholls, J. T.; Frost, J. E. F.; Linfield, E. H.; Pepper, M.; Ford, C. J. B.; Ritchie, D. A.; Jones, G. A. C.; Kogan, E.; Kaveh, M. *J. Phys.: Condens. Matter* **1994,** *6*, 4763.

19. Wang, T.; Clark, K. P.; Spencer, G. F.; Mack, A. M.; Kirk, W. P. *Phys. Rev. Lett.* **1994,** *72*, 709-712.

20. Wei, H. P.; Tsui, D. C.; Pruisken, A. M. M. *Phys. Rev. B* **1986,** *33*, 1488-1491.

21. Burgess, C. P.; Dolan, B. P. *Phys. Rev. B* **2007,** *76*, 113406.

22. Khemelinskii, D. E. *Pis'ma Zh. Eksp. Teor. Fiz.* **1983,** *38*, 454 [*JETP Lett.* **1983,** *38*, 552].

23. Laughlin, R. B. *Phys. Rev. Lett.* **1984,** *52*, 2304-2304.

24. Kivelson, S.; Lee, D.-H.; Zhang, S.-C. *Phys. Rev. B* **1992,** *46*, 2223-2238.

25. Ortmann, F.; Roche, S. *Phys. Rev. Lett.* **2013,** *110*, 086602.

26. Zhang, L.; Zhang, Y.; Khodas, M.; Valla, T.; Zaliznyak, I. A. *Phys. Rev. Lett.* **2010,** *105*, 046804.

27. Giesbers, A. J. M.; Zeitler, U.; Katsnelson, M. I.; Ponomarenko, L. A.; Mohiuddin, T. M.; Maan, J. C. *Phys. Rev. Lett.* **2007,** *99*, 206803.

28. Peng, X.; Ahuja, R. *Nano Lett.* **2008,** *8*, 4464-4468.

Figure Captions

Fig. 1 $\rho_{xx}$ and $\rho_{xy}$ at different temperatures $T$ for (a) EG1, (b) EG2, and (c) EG3. The vertical arrows indicate the temperature increase: $T$ = 2.52 K, 3.50 K, 4.25 K, 5.50 K, 7.00 K, 8.50 K, and 10.0 K for EG1; $T$ = 2.60 K, 3.54 K, 4.55 K, 5.56 K, and 7.00 K for EG2; $T$ = 4.45 K, 7 K, 10 K, 15 K, and 25 K for EG3.

Fig. 2 The directly converted conductivities, $\sigma_{xx}$ and $\sigma_{xy}$, at different $T$ for (a) EG1, (b) EG2, and (c) EG3. The vertical arrows indicate the temperature increase. The temperature points are the same as those given in the caption of Fig. 1 for each sample.

Fig. 3 Conductivity $\sigma_{xx}$ plotted against $\sigma_{xy}$ for (a) EG1, (b) EG2 and (c) EG3. The dotted curves denote the theoretical prediction of semicircle $\sigma_{xx}$-$\sigma_{xy}$ relation for the 0-2 transition. Each group of triangle markers connected by dashed lines denotes the data for the same magnetic field. The arrows indicate the flow line to the low temperature extreme at fixed magnetic fields. The black ones correspond to the flow at the observed crossing point $B_c^\sigma$ in $\sigma_{xy}$.

Fig. 4 Schematic diagrams illustrating the floating up of the electron zeroth ($0^{th}$) LL with decreasing $B$ in EG. In contrast, the $0^{th}$ LL is $B$-independent in PG.

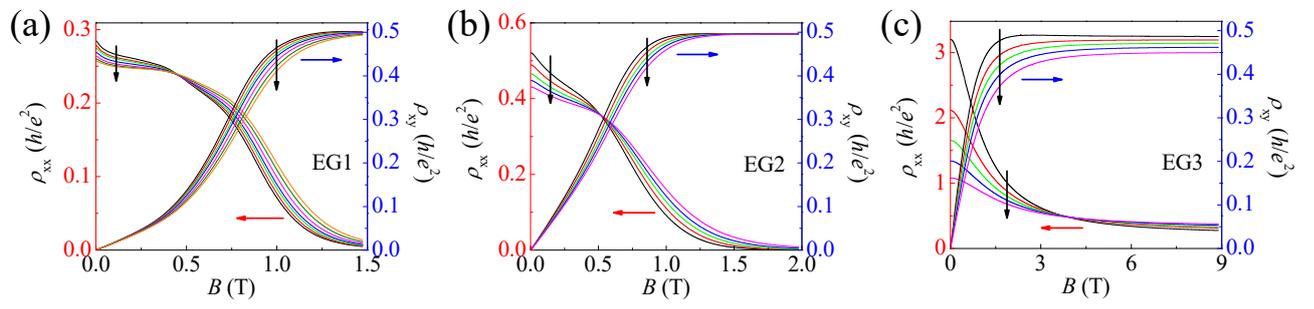

Figure 1

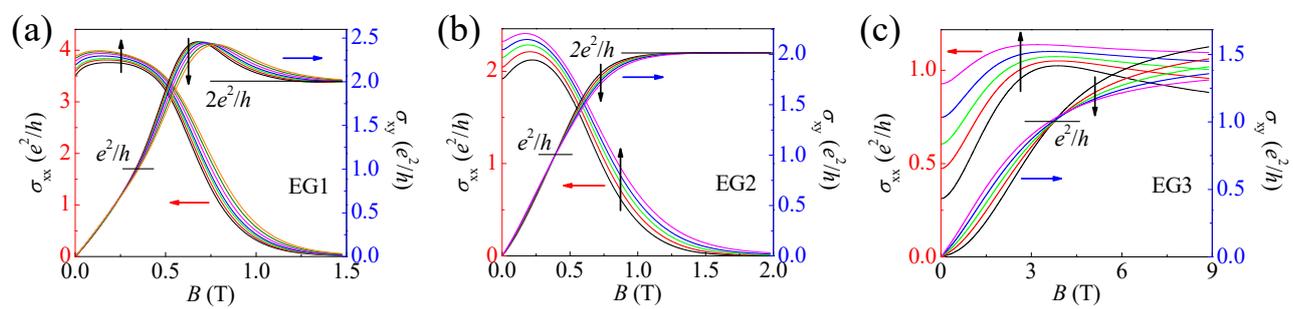

Figure 2

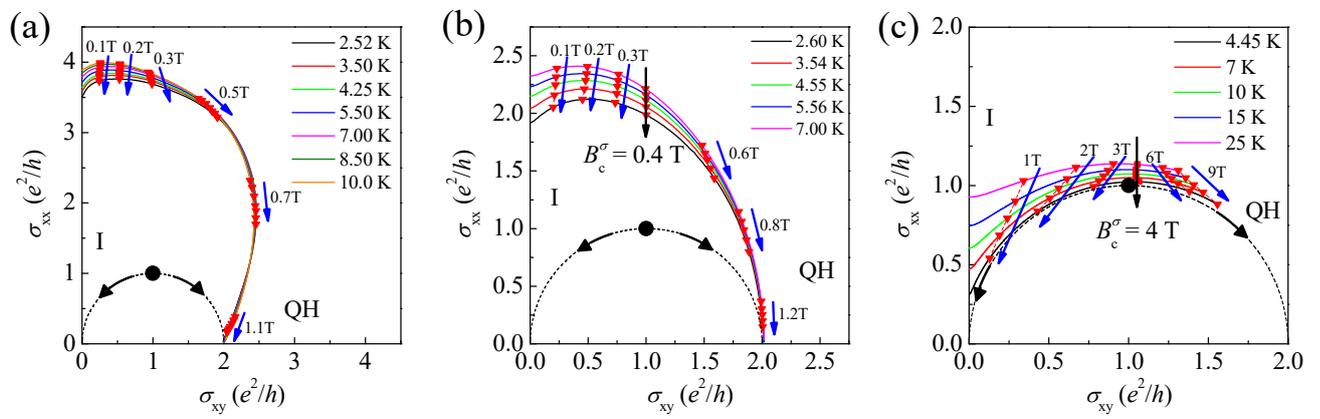

Figure 3

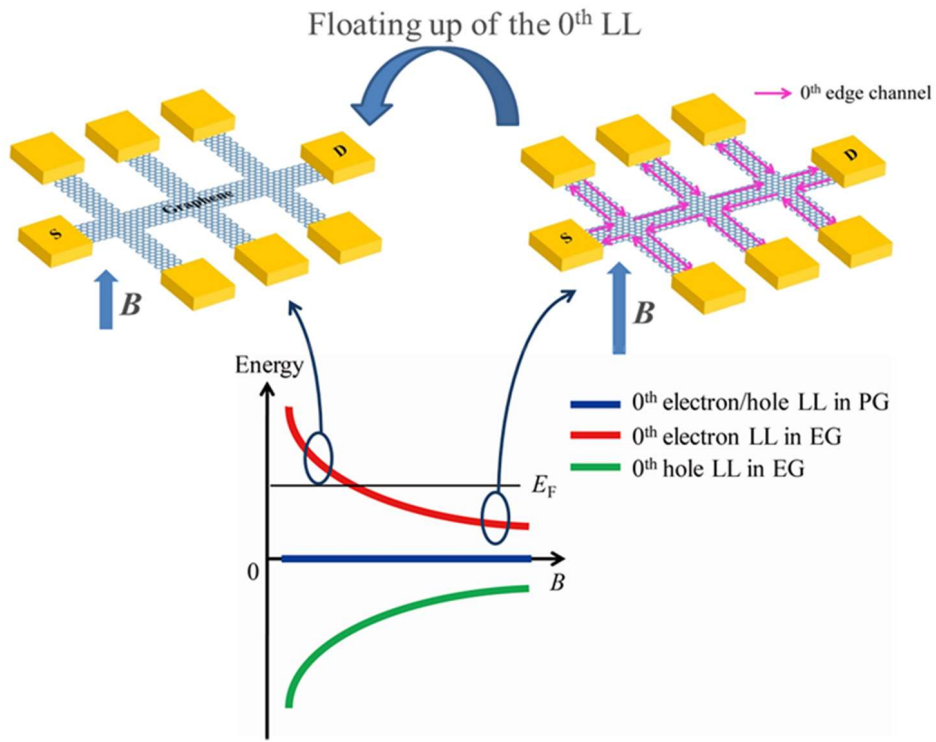

Figure 4

# Floating up of the zero-energy Landau level in monolayer epitaxial graphene

Lung-I Huang[1,2], Yanfei Yang[1,3], Randolph E. Elmquist[1],
Shun-Tsung Lo[4*], Fan-Hung Liu[4], and Chi-Te Liang[2,4*]

[1]*National Institute of Standards and Technology (NIST), Gaithersburg, MD 20899, USA*
[2]*Department of Physics, National Taiwan University, Taipei 106, Taiwan*
[3]*Department of Physics, Georgetown University, Washington, DC 20057, USA*
[4]*Graduate Institute of Applied Physics, National Taiwan University, Taipei 106, Taiwan*

*shuntsunglo@mail.ncku.edu.tw and ctliang@phys.ntu.edu.tw

**Sample preparation and measurements**

Epitaxial graphene (EG) is formed after decomposition and Si sublimation on the surface of SiC at high temperatures. Angle-resolved photoelectron spectroscopy shows that newly-grown samples measured *in situ* have carrier concentrations $n \approx 10^{13}$ cm$^{-2}$, ascribed to charge-transfer from an insulating graphene-like buffer layer that is covalently bonded to the SiC substrate [1]. In order to study the electronic transport with $|n| < 10^{12}$ cm$^{-2}$, electrostatic [2, 3] or photochemical [4] gating through an insulating dielectric, molecular doping [5] directly on the EG surface, or atomic intercalation [1, 6] beneath the buffer layer have been used to modify the carrier concentration. In order to achieve low density EG, Our EG devices were fabricated utilizing a clean lithography process [7] that leaves the surface free of resist residues. After this fabrication process doping occurs due to or initiated by chemical etching of the protective layer and exposure to air, producing typical carrier densities of order $n \approx 10^{11}$ cm$^{-2}$. The devices can be cycled to higher or lower carrier density repeatedly by annealing at 70 °C to 150 °C or by air exposure, implicating oxygen and water molecules from the air as the source of p-type molecular doping [8, 9].

Longitudinal resistivity $\rho_{xx}$ was obtained by averaging the data from both sides of the conducting channel [voltage probes 1, 3 and voltage probes 1* and 3*] and Hall resistivity $\rho_{xy}$ was measured across the central pair [2 and 2*] of device contacts [Fig. S1]. In graphene as well as in heterostructures, low carrier concentrations are often associated with percolating current paths that mix $\rho_{xx}$ with $\rho_{xy}$. Data measured at both directions of the magnetic field were combined based on the recognized symmetries of the resistivity components to eliminate this mixing [10], which is strong in highly disordered samples for large values of $\rho_{xx}$.

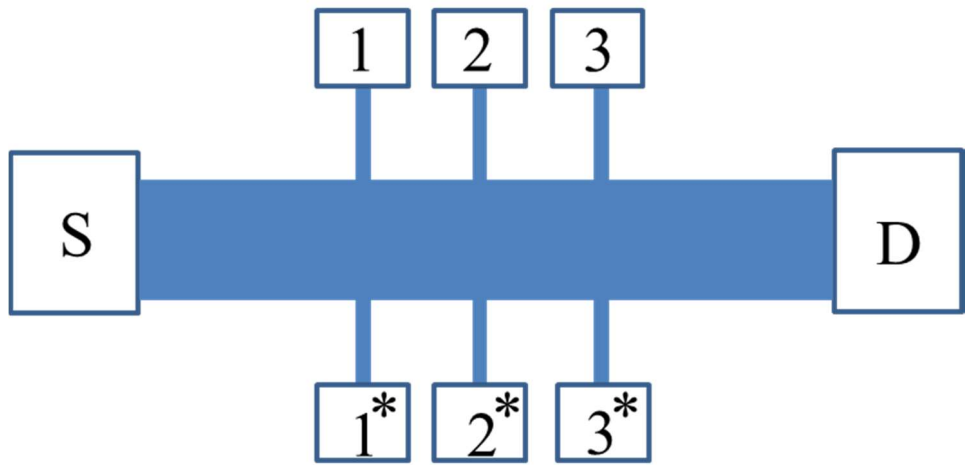

Figure S1 Schematic diagram showing a typical monolayer epitaxial graphene (EG) sample. S and D correspond to source and drain contacts. 1, 2, 3, $1^*$, $2^*$ and $3^*$ are voltage probes. Channel dimensions, which are the same for all devices studied, are $L = 0.6$ mm, $W = 0.1$ mm, with voltage contacts spaced 0.1 mm apart along both sides of the device.

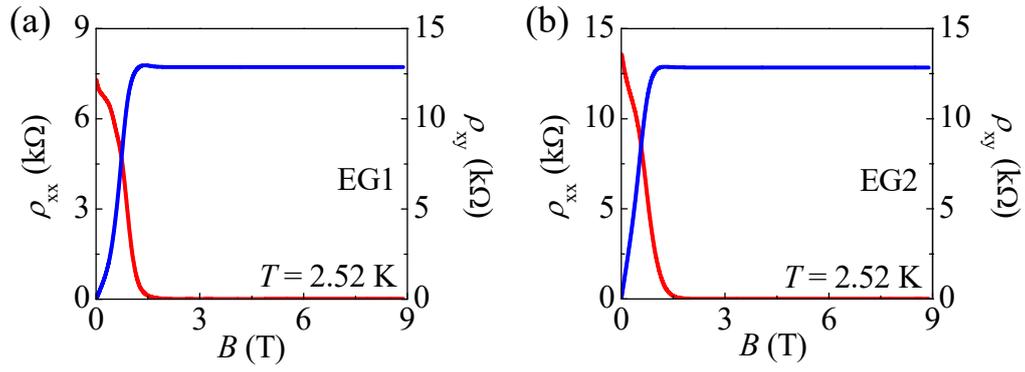

Figure S2. Resistivity values $\rho_{xx}(B)$ and $\rho_{xy}(B)$ of samples (a) EG1 and (b) EG2 for $0 < B < 9$ T.

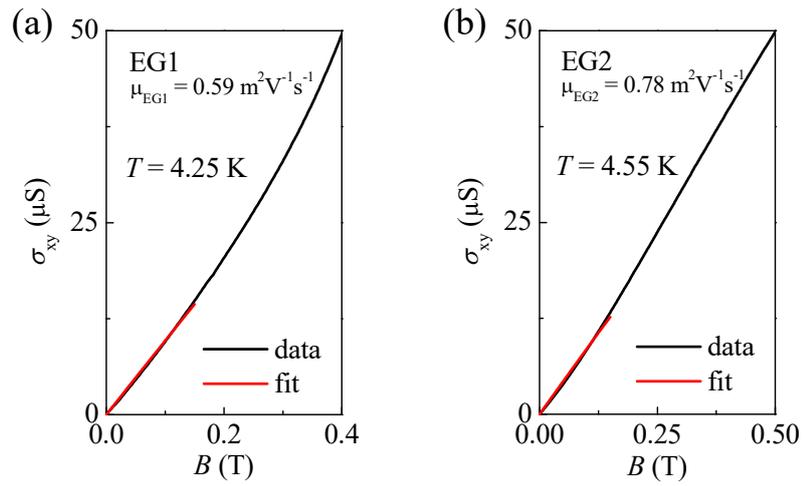

Figure S3 Determination of the mobility $\mu$ for samples (a) EG1 and (b) EG2 by fitting the measured $\sigma_{xy}$ to $ne\mu^2 B/(1+(\mu B)^2)$ over the range of $0 < B < 0.15$ T.

**Weak localization and electron-electron interactions in our devices**

In the weakly disordered regime, that is, the conductivity higher than $e^2/\pi h$, weak localization (WL) and electron-electron interaction (EEI) have significant contributions to the transport at low $B$ in disordered graphene devices and may influence [11] the observed I-QH transitions [12-16]. The WL term modifies $\rho_{xx}$ without affecting $\rho_{xy}$. The diffusive EEI has effects on both $\rho_{xx}$ and $\rho_{xy}$. To investigate the observed I-QH transition, we have isolated the EEI contribution from the WL one following Ref. [17]. The EEI correction to the Drude conductivity [17] is given by

$$\delta\sigma_{xx}^{ee} = -K_{ee} G_0 \ln(\frac{\hbar}{k_B T \tau}), \qquad (1)$$

where $K_{ee}$ is an interaction parameter dependent on the type of sample and $\tau$ is the scattering time. This term gives a $\ln T$ dependence to both $\sigma_{xx}$ and to the Hall coefficient $R_H \equiv \delta\rho_{xy}(B, T)/\delta B$. The $\ln T$ dependence of $R_H$ is shown in Fig. S4 (a).

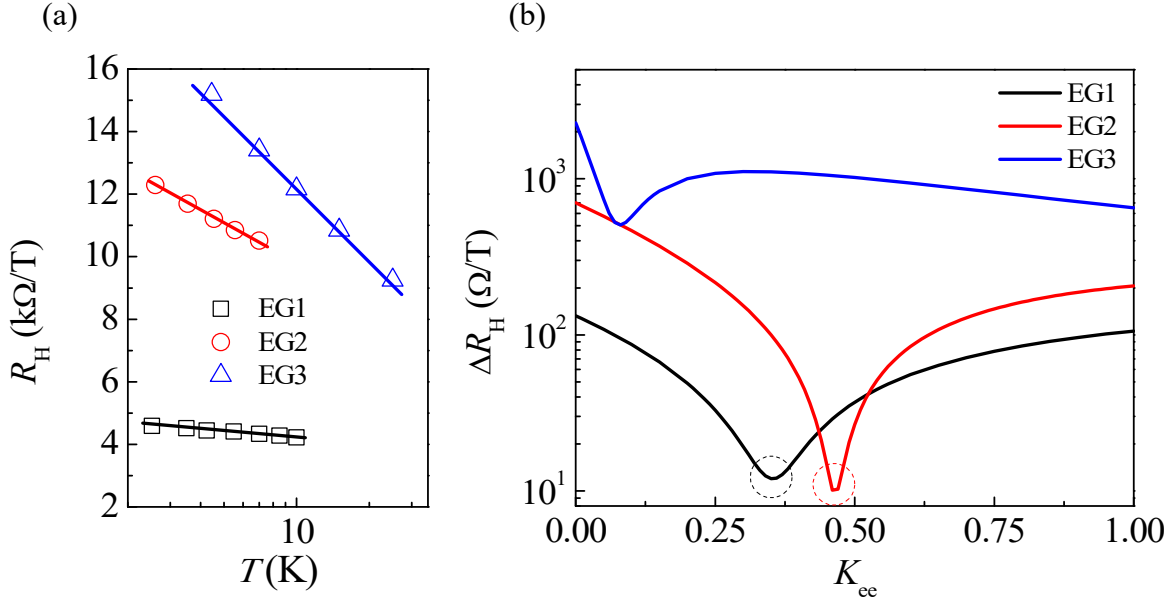

Fig. S4 (a) Uncorrected Hall slope $R_H \equiv \delta\rho_{xy}(B, T)/\delta B$ as a function of $T$. (b) Standard deviation of the corrected Hall slope at different $T$, $\Delta R_H = \sqrt{\frac{1}{N-1}\sum_i (R_H^i - \overline{R_H})^2}$ (where i runs over the measured temperature points), plotted against the interaction parameter $K_{ee}$. $\Delta R_H$ of the uncorrected data in (a) for each sample corresponds to $\triangle R_H(K_{ee} = 0)$ in (b).

According to Eq. (1), matrix inversion of the conductivity tensor shows that $\rho_{xx}(B,T)$ takes a parabolic form [18],

$$\rho_{xx} \approx \frac{1}{\sigma_D} - \frac{1}{\sigma_D^2}(1-\mu^2 B^2)\delta\sigma_{xx}^{ee}(T), \tag{2}$$

for $\delta\sigma_{xx}^{ee} \ll \sigma_D$, where $\mu$ is the mobility, $\sigma_D$ is the Drude conductivity and $\mu$ is the mobility. In addition, the EEI term gives a correction to the Hall coefficient $R_H \equiv \delta\rho_{xy}(B, T)/\delta B$ following $\delta R_H / R_H^0 = -2\delta\sigma_{ee}/\sigma_D$, where $R_H^0$ denotes the classical value of $R_H$ [17]. The ln$T$ dependence of $R_H$ is observed in Fig. S4(a), suggesting the influence of electron-electron interactions on the low-field insulating behavior.

Relevant to the data analysis, Eq. (2) indicates a $T$-independent point in $\rho_{xx}$ at $\mu B = 1$. To clarify this its relation with the observed crossing issue, we remove the correction the contribution of EEI as described by Eq. (2) to $\rho_{xx}$ at low $B$ [18] and estimate the EEI strength following Ref. [17]. The correction $\delta\sigma_{xx}^{ee}$ described by Eq. (2) is subtracted from the measured $\sigma_{xx}$ for with $0 \leq K_{ee} \leq 1$. By inverting the resulting conductivity tensor, we obtain a new corrected set of $\rho_{xx}$ and $\rho_{xy}$. The optimum $K_{ee}$ is identified when the standard deviation of the corrected $R_H$ values at different $T$ in Fig. S4(b) reaches its minimum. As shown in Figs. S5(a) and S5(c), for EG1 and EG2 the correction removal process renders the corrected $\rho_{xy}$ insensitive to the change in $T$ at low fields and the slope corresponds to $R_H^0$ without suffering from EEI. Most disordered device does not produce an optimum $K_{ee}$ with reasonable confidence, and only a weak minimum (EG3) is obtained by this procedure. The $T$-independent points in $\rho_{xx}(B, T)$ survive in the corrected data for EG1 and EG2 and occur at only slightly lower crossing fields $B_c^\rho$ after the correction [Figs. S5(a) and S5(b)]. The remaining $T$ and $B$ dependence of $\rho_{xx}$ is attributed to WL effect (Supplementary Fig. S5), suggesting that the transition in EG1 and EG2 represents the crossover from WL to the $\nu = 2$ quantum Hall state. However, stronger disorder in EG3 whose low-$T$ conductivity is lower than $e^2/\pi h$ makes the correction descriptions invalid.

**Remove the corrections due to electron-electron interactions**

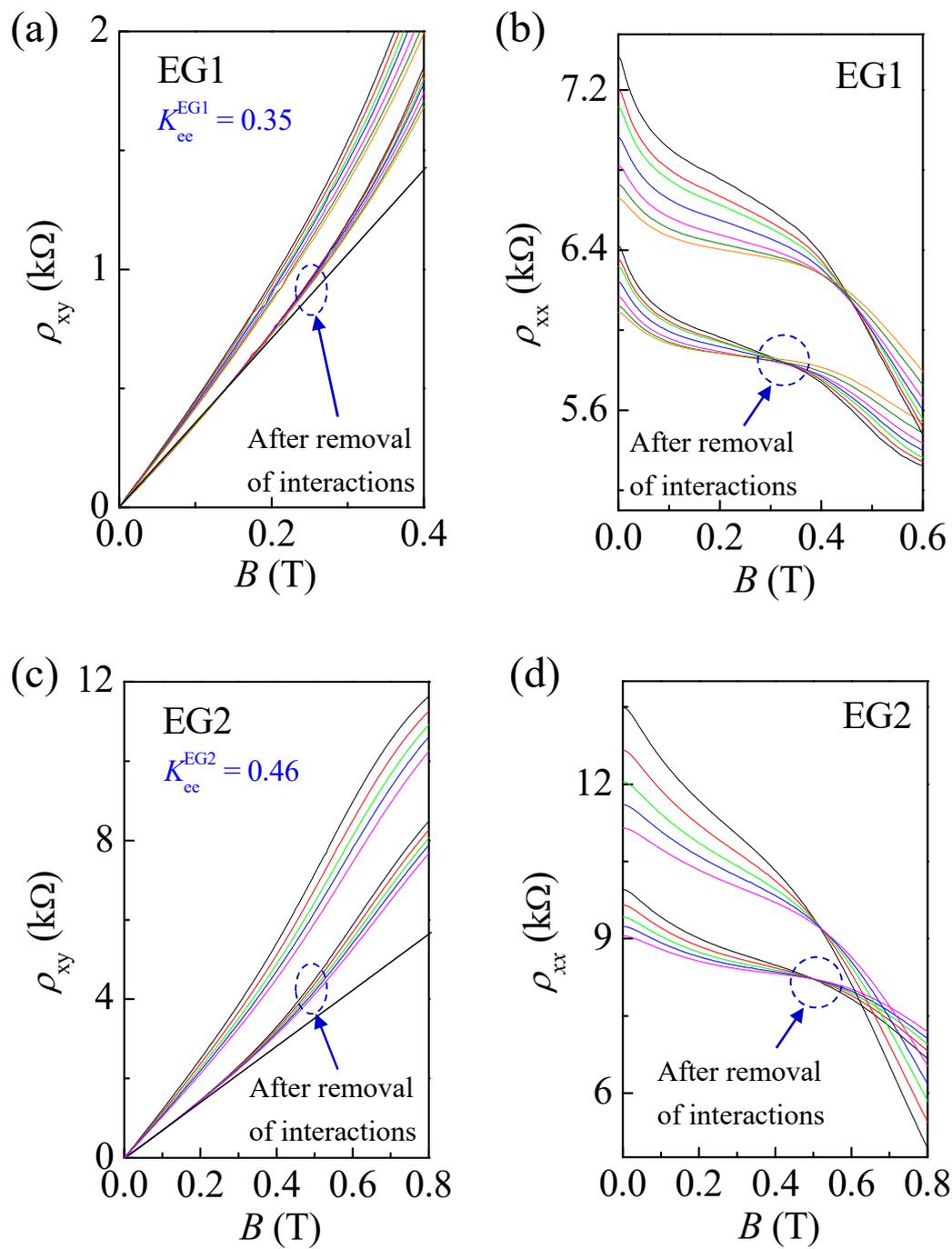

Fig. S5 Comparison of $T$-dependent resistivities for samples (a, b) EG1 and (c, d) EG2 before and after removal of interactions. The temperature ranges are the same as those given in the caption of Fig. 1.

**Weak localization**

Our experimental results can be fitted to the theoretical work of McCann *et al.* [19] as shown in Fig. S6 (a) and (b). We note that the WL effect contributes to a shift in $\sigma_{xx}$ proportional to $\ln(\tau_\phi/\tau)$, where $\tau_\phi$ is the phase relaxation time and approximately proportional to $T^{-1}$ as shown in Fig. S6 (c); however, WL produces no contribution to Hall coefficient.

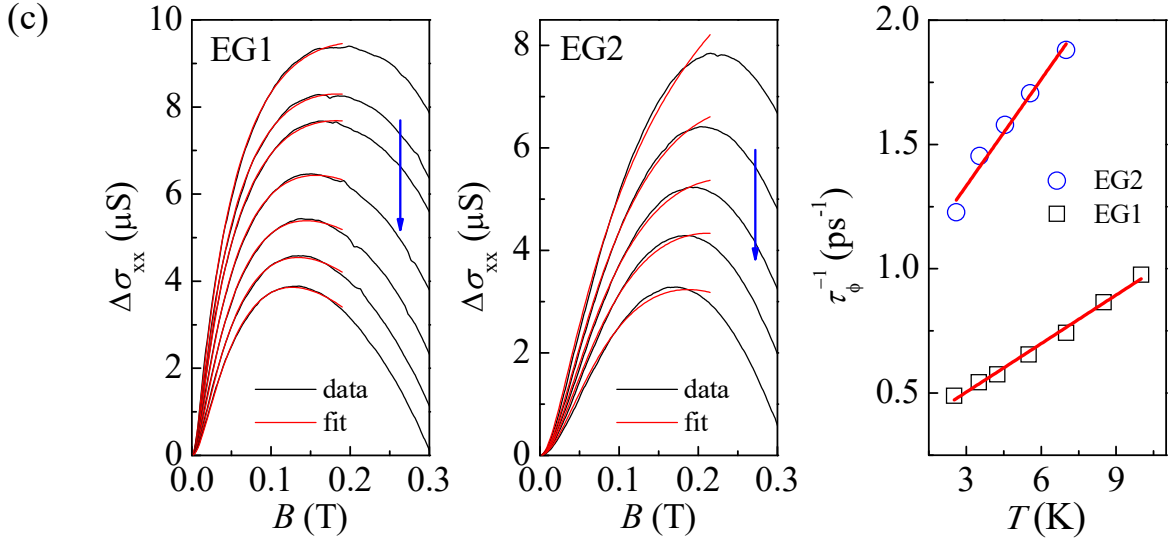

(c) ... $\equiv \sigma_{xx}(B) - \sigma_{xx}(B = 0)$ to the model developed ... EG1 and (b) EG2. The arrows indicate the ... rate $\tau_\phi^{-1}$ obtained from the fits as a function

## Scaling of the Hall conductivity

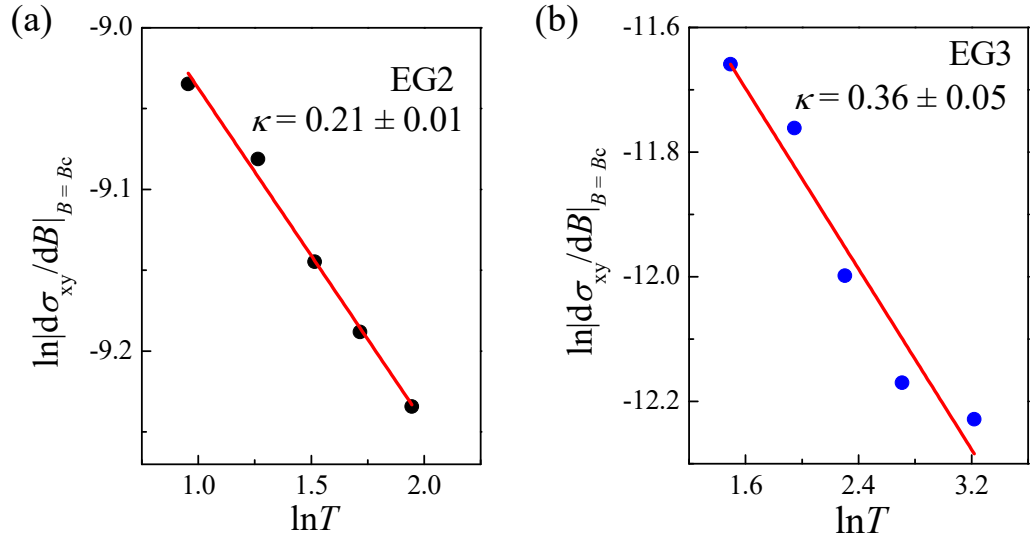

Figure S7 Fit of the slope of the transverse conductivity $d\sigma_{xy}/dB$ at the critical field $B_c^\sigma$ to the power-law dependence on temperature $T$ with an exponent $\kappa$ for EG2 and EG3.

Table S1 Physical quantities of each EG sample.

| Sample | Type | density (m$^{-2}$) | $K_{ee}$ | $\mu$ (m$^2$V$^{-1}$s$^{-1}$) | $\tau$ (fs) | $\Gamma$ (meV) | $\mu B_c^\rho$ |
|---|---|---|---|---|---|---|---|
| EG1 | n | $1.75 \times 10^{15}$ | 0.35 | 0.59 | 29 | 23 | 0.27 |
| EG2 | p | $8.83 \times 10^{14}$ | 0.46 | 0.78 | 27 | 24 | 0.41 |
| EG3 | n | $5.76 \times 10^{14}$ | – | 0.31 | 9 | 76 | 1.21 |

**References**

1. J. Ristein, S. Mammadov, and Th. Seyller, *Phys. Rev. Lett.* **108**, 246104 (2012).
2. D. B. Farmer *et al.*, *Phys. Rev. B* **84**, 205417 (2011).
3. T. Shen *et al.*, *J. Appl. Phys.* **111**, 013716 (2012).
4. S. Lara-Avila *et al.*, *Adv. Mater.* **23**, 878 (2011).
5. C. Coletti *et al.*, *Phys. Rev. B* **81**, 235401 (2010).
6. P. L. Levesque *et al.*, *Nano Lett.* **11**, 132 (2011).
7. Y. Yang *et al.*, *Small* **11**, 90 (2015).
8. S. Ryu *et al.*, *Nano Lett.* **10**, 4944 (2010).
9. Z. H. Ni *et al.*, *J. Raman Spectros.* **41**, 479 (2010).
10. M. Hilke *et al.*, *Nature* **395**, 675 (1998).
11. M. Y. Simmons *et al.*, *Phys. Rev. Lett.* **84**, 2489 (2000).
12. S. H. Song *et al.*, *Phys. Rev. Lett.* **78**, 2200 (1997).
13. H. W. Jiang *et al.*, *Phys. Rev. Lett.* **71**, 1439 (1993).
14. R. J. F. Hughes *et al.*, *J. Phys.: Condens. Matter* **6**, 4763 (1994).
15. T. Wang *et al., Phys. Rev. Lett.* **72**, 709 (1994).
16. E. Pallecchi *et al.*, *Sci. Rep.* **3**, 1791 (2013).
17. K. E. J. Goh, M. Y. Simmons and A. R. Hamilton, *Phys. Rev. B* **77**, 235410 (2008).
18. G. M. Minkov *et al.*, *Phys. Rev. B* **67**, 205306 (2003).
19. E. McCann *et al., Phys. Rev. Lett.* **97**, 146805 (2006).